# AN EFFICIENT INDOOR NAVIGATION TECHNIQUE TO FIND OPTIMAL ROUTE FOR BLINDS USING QR CODES

Affan Idrees, Zahid Iqbal, Maria Ishfaq

Department of Computer Science, University of Gujrat, Pakistan.

[10070619-017, zahid.iqbal, 10070619-004] @uog.edu.pk

## ABSTRACT

Blind navigation is an accessibility application that enables blind to use an android Smartphone in an easy way for indoor navigation with instructions in audio form. We have proposed a prototype which is an indoor navigation application for blinds that uses QR codes. It is developed for android Smart phones and does not require any additional hardware for navigation. It provides automatic navigational assistance on pre-defined paths for blind. QR codes are placed on the floor sections after specific distance that acts as an input for current location detection and navigation. Whenever a QR code is scanned it provides the user with the information of the current location and asks the user to select the destination and then offers optimal and shortest path using path finding algorithms. During navigation whenever the deviation from the proposed path is detected it prompts the user and guides back to the right path by comparing the current path with the generated path. All of the instructions throughout the application are provided in audio form to the user. The interface of the application is well built for blinds which makes the smart phones user-friendly and useable for blind people. The user interacts with the application through a specific set of user-friendly gestures for specific inputs and operations. At the end, we have performed comparison between different state of art approaches and concluded that our approach is more user friendly, cost effective and produced more accurate results.

## 1. INTRODUCTION

In contemporary world technology has become a necessity of every human being, each one of us has to interact with technological objects in some way for a lot of different purposes. According to the "Visual Impairment and Blindness Factor Sheet" last updated in August 2014 by World Health Organization about 285 million people are visually impaired across the world, among which 39 million are blind whereas 246 million have low vision. Most of the visually impaired people i.e. approximately 90%; belong to the developing countries. The advancement in the field of technology is emerging every minute but there are still some problems faced by visually impaired persons in using modern technologies. Mostly blind people do not get involved in social or physical activities. They are even reluctant in visiting new places and require assistance from someone else in performing several activities. They usually use a white cane or in some cases trained dogs for moving around. Due to the absence of vision blinds are wholly or partially dependent. Evolution of technology over the past few decades has introduced many solutions for the problems faced by the people in various fields. Technology has also provided many solutions gadgets and applications [1]. Laser sensors integrated in the helmets for hurdle detection that a blind can wear, high definition image processing cameras with embedded circuits trained along some particular path that work by processing and comparing the images of the path being followed with the already saved ones [2], line follower applications [3] and sensors integrated within the cane [4] are among some solutions presented recently by the technology for the blinds. Apart from it, solutions to many other problems have also been provided by Artificial Intelligence in the form of; Swarm optimizer for locating and tracking multiple optima [5], hybrid approaches used for UCTP [6], machine learning techniques for stock price prediction [7] and systematic study on OC techniques [8]. The solution proposed by our research aims to offer eyes-free technological support for blind people to walk around and visit specified locations with proper guidance along the generated path indoors. Our solution is designed by keeping in mind the difficulties faced by visually impaired persons.

## 2. RELATED WORK

Technology has been helping disabled people in many ways in performing different activities e.g. automatic wheel chairs for people with mobility impairment, speech output systems for people with low vision and specially designed visual output for people with hearing impairment [1]. Now with the help of technology blind persons have ease of mobility, shopping, recognizing their objects, and detecting obstacles along their path in their daily routine.

Vinod Pathangay in [2] suggests an Indoor navigation technique based on image processing in which assistive camera is mounted on walker. With the help of image processing algorithm, based upon temporal matching of graphics data obtained at run-time with already saved data during training period, correctness of path is measured and alarm is generated in case of deviation. However sometimes there is greater false alarm ratio for same paths since slight change of angle of camera can result in greater dissimilarity ratio.

Another comparatively simple and quicker indoor navigation system is proposed in [3], ARIANNA (project name) functions by recognizing specific colored tapes stick on the floor. User has to place a finger on the screen and move the device in the horizontal direction device will produce the vibration when finger touches the lane. QR codes are stick near the POIs (points of interest) that contain a URL which is used for fetching the instructions from path server using a Wi-Fi connection.

Blind Shopping application, an idea presented in [4] offers shopping for blinds with features like eyes-free product search, selection and navigation with-in the store that will already contain QR codes for identification of product categories. It also suggests integration of RFID tags on the floor and the user carries a RFID reader i.e. a portable device that senses those tags and calculates the current position of the user. Since it fetches the data from an online database so a working Wi-Fi connection is required for its smooth working which shows its dependence upon internet connection.

Slim Kammoun in [9] suggests based on his research and interviews held with blinds that blind people prefer optimal path (one with the low traffic) selection rather than the shortest path from the current location to the destination position. The route was selected using Geographical Navigation System and Dijkstra's Algorithm.

In [10] use of image processing algorithms in a different way is presented as a solution for indoor navigation for blinds. The project is divided in two major portions, first part comprises of detection of the current location accurately that is based upon matching appearances and second part consists of guiding along the path that uses a path generation algorithm.

In [11] two types of algorithms are discussed, first an algorithm is described that plan the full route before moving and then the second algorithms that plans path while navigating for blinds. For optimal path finding algorithms that plan full route before moving are suggested i.e. Best First Search, Depth First Search, Dijkstra's and A*. Global space-search and optimal algorithm in path finding is A* a combination of Dijkstra's and Best first search.

In [12] optimization of A* is discussed. In large space algorithm has to process more which results in higher time complexity. Hierarchical path finding A* is a solution of large space path finding problem. It reduces the complexity hierarchal of search space.

Adam Herout discuss in [13] fast detection and recognition of QR codes from high resolution images that not only contain QR codes but may contain text and other images. Algorithm successfully localized 95% of all QR codes. The required run-time was under 250ms in average for the tested images (up to 15MPix large).

Yue Liu and Mingjun Liu discuss in [14] & [15] benefits of using QR code. QR code is a typical matrix two-dimensional barcode. QR codes are efficient than the barcodes on the basis of the following reasons: 1) High capacity encoding of the data, 2) High speed reading, 3) Readable from any direction from 360 degrees.

## 3. PROPOSED SYSTEM

Indoor navigation has always been a challenge; several techniques have been proposed to solve this problem. For blinds it has to be more precise than for the sighted people. Our study proposes implementation of an optimal path finding algorithm for a pre-defined scenario. The Proposed system requires a custom scenario with QR codes attached to the floor sections in the form of horizontal strips and an android smart phone with integrated webcam having OS version 2.2(Froyo) and higher. To minimize th e chances of skipping a QR, we decided to place horizontal strips of QR across full width of the path for error free scanning of QR codes (fig 1).

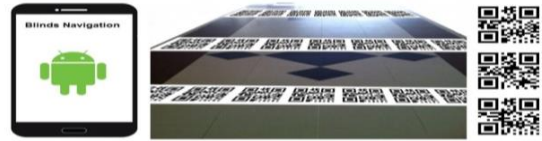

Fig 1: Android Smartphone, QR Strips in corridor, QR codes (Left to right)

### 3.1 Scenario Description

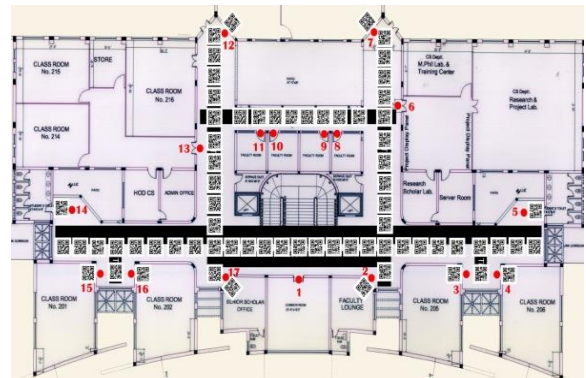

Fig 2: Faculty of Computing and Information Technology Block, 2nd Floor

Demonstration area of the application is Faculty of Computing and Information Technology Block, University of Gujrat. There are 17 possible destination locations (shown with red dots) in the main corridor. All the location points can be source and intermediate nodes. QR code strips are pasted on the floor section horizontally after regular intervals.

### 3.2 Description of Approach

Our proposed system for blind navigation consists of following modules shown in fig 3. These modules interact with each other and make a perfect accessible interface for blinds designed in such a way that enhances the usability of an android app for blinds. Our app can be divided into six major modules. These modules interact with each other and perform specific functionalities to support smooth guidance along the generated path.

First one is the QR Scanner, which scans QR codes and provides the data encoded in the QR codes to the application. This data is used for fetching the corresponding instructions from the database. The library that we are using for QR code scanning is "ZXing" and it is a better option for scanning large QR codes.

As described in the 1st phase of diagram, whenever a QR code is scanned an instruction is fetched from the SQLite database (the second module) which carries the information about all the nodes of a particular scenario. After the instruction is being fetched, it is being passed on to the output screen and provided to the user in the form of audio/speech.

The third module is Text2Speech which is a key for providing an accessible interface to the blinds along with specific gestures. All of the instructions that are provided to the user during navigation are in audio form. After providing user with the information about current location, the application switches to the destination menu (Phase 2) that includes all of the possible destinations. Once the destination has been selected, both the current location and destination values are passed on to the Path generation module.

At first the user is asked about which path user wants to generate i.e. either the shortest one or the optimal one. Shortest Path here is generated on the basis of the distance from current location to the destination location whereas optimal path is generated on the basis of the number of turns from the current location to the destination location. Once the path is generated it gets stored in the next module i.e. Trip Manager. Trip Manager is responsible for the successful navigation of the user along the generated path which is accompanied by another module i.e.

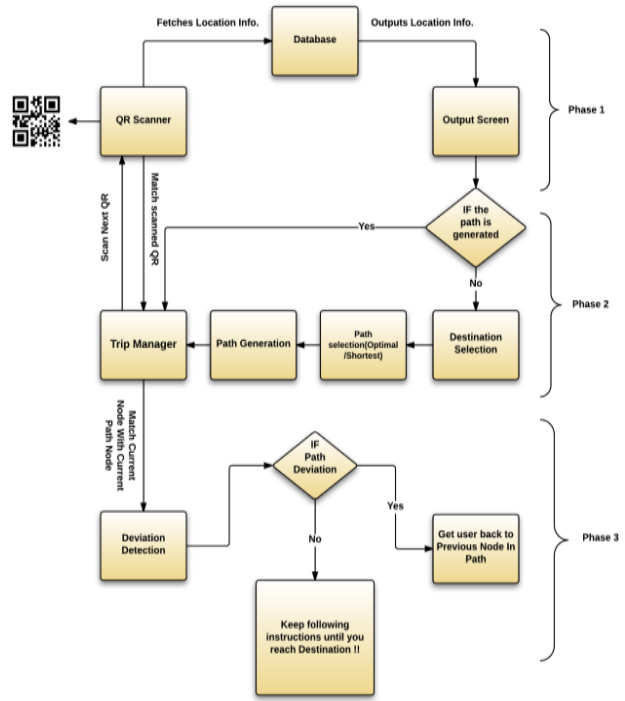

Fig 3: Operational Diagram

Deviation Detection (6th Modules). Trip Manager ensures the smooth navigation along the generated path unless the user gets deviated from that path. If the user gets deviated, this deviation is checked with the help of current node comparison with the current node to be visited according to the generated path. If the nodes do not match, deviation detection module prompts the user about deviation from the correct path. The technique that we have adopted to get the user back on correct path in deviation detection and correction is based upon guiding the user back to previous correct node visited by the user according to the path. In this way the application not only provides the user with the path of his choice but also keeps checking the accuracy of the path.

## 4. Results and Discussion

To test our proposed system we first placed the QR strips on the floor section of the Faculty of Computing & Information Technology, University of Gujrat. Each QR strip was at a distance of approximately four steps from each other. The interface of the application is well built for blinds and special gestures are used keeping in mind the difficulties faced by them in using touch screen smart phones.

Description of special gestures used in our application is as follows:

- Single tap gesture is used for repeating current selection or instruction in audio form
- Double tap gesture is for confirmation of current selection in the whole app
- Swipe Right/Left gesture for navigating between the activities
- Swipe Up/Down gesture for navigating between the list items

We performed test of our QR code scanning module under different light conditions, with different sizes of QR codes and different ratios of blurriness of QR Codes. QR codes were easily detected under low light conditions and up to 60% of blurriness ratio.

| Blurriness Ratio | Scan Successful |
|---|---|
| 10% | Yes |
| 20% | Yes |
| 30% | Yes |
| 40% | Yes |
| 50% | Yes |
| 60% | Yes |
| 70% | Yes (With Delay) |
| 80% | No |
| 90% | No |

Table 1: QR Code Detection

First we tested the application by scanning the first QR code in front of the Location # 1 that provided the information about our current location and then selected Location # 10 as a destination. Then we selected optimal path and the app then provided the instruction for next QR along the generated path. Whenever a QR was scanned an instruction was fetched which guided to the user in audio form. Upon reaching the selected destination app provided "Destination Reached" message and then asks the user to either go back to the location from where he started the trip or start a new trip from his present location. In this way the app successfully guided us along the generated path.

In the second case, we tested the deviation detection module by scanning the node that was not in the generated path. In this case the app prompted the user about path deviation through vibration and a message in the audio form. Deviation correction was done by guiding the user to the previous correct node visited. After reaching the correct node the user was guided along the generated path to the destination.

We have analyzed some indoor navigation system on the basis of seven core features:

| Project Names / Features | [2] | [3] | [4] | [5] | Blinds Navigation |
|---|---|---|---|---|---|
| Current Location Detection | No | Yes | Partial (Near POIs) | Yes | Yes (On Every QR) |
| Cost effective | No | Yes | Yes | Yes | Yes |
| Accessible Interface | No | Yes | Yes | No | Yes |
| Optimal Path Finding | No | No | No | Yes | Yes |
| Standalone Application | Yes | No | No | Yes | Yes |
| Deviation Detection | Yes | No | Yes | No | Yes |
| Response Time | High | High | High | High | Low |

Table 2: Comparison of different indoor navigation systems

For generation of optimal path based upon our study we found A* algorithm to be best for this purpose. A* requires heuristic data for optimal path generation. For this purpose we enhanced heuristic value calculation method to provide the blinds with the best available path from the user's point of view. The algorithm we have used for this purpose finds all possible paths from current location to destination and finds its heuristics either on the basis of turn i.e. optimal path or on the basis of distance i.e. shortest path.

There are many other path finding algorithms i.e. Breadth First Search, Depth First Search, Greedy Search. Breadth First search is complete but does not offer optimal path since it stops greater memory requirements. Depth First Search uses less memory than BFS but it is neither complete nor optimal. Since greedy search picks the node with minimum edge distance from the available nodes thus it is neither complete nor optimal.

For example, for optimal path, if user has selected Location "1" as source point and Location "13" as destination point for optimal path. There will be two possible paths in this case from selected source point to destination shown in the figures below.

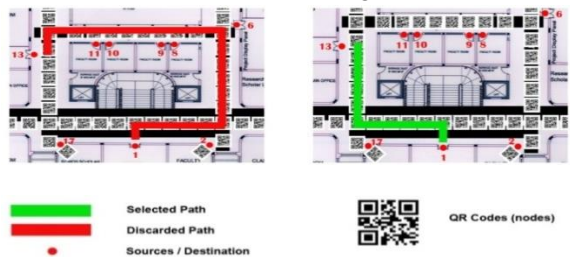

Fig 4: Optimal Path Selection

We have selected optimal path here, on the basis of heuristic value, red path is discarded by algorithm since its heuristic value is greater than the heuristic value of green.

## 4. Conclusion and Future Work

The subject area of our study was based specially on smooth navigation for blinds indoors. We aimed at increasing the usability of android smart phone for blinds and use it for their navigation. We made our research and developmental study to deploy "artificially intelligent path finding technique" for android Smartphone.

There are a number of ways this research work can be enhanced to provide additional features to the system:

- Map data can be added for several location based on user's interests.
- A module can be added for effective hurdle detection along the generated paths.
- Hurdle detection module can be enhanced by adding pre-hurdle alert for saved routes.
- Object recognition module can also be added to the system that will help the blind in recognizing their objects.
- Online support for downloading the map data for new scenarios can be added.
- With the help of GPS, this system can be enhanced to be used for both indoors and outdoors.
- Multiple language support can also be added to the system.
- System can developed to receive instructions in audio form as an input.